\documentclass[aps,pre,twocolumn,groupedaddress]{revtex4-2}
\usepackage{graphicx}
\usepackage{dcolumn}
\usepackage{bm}
\usepackage{amsmath, amssymb,bbm}
\usepackage{xcolor}

\newcommand{\C}{\mathbb{C}}
\newcommand{\R}{\mathbb{R}}
\newcommand{\N}{\mathbb{N}}

\newcommand{\bea}{\begin{eqnarray}}
\newcommand{\eea}{\end{eqnarray}}
\newcommand{\beast}{\begin{eqnarray*}}
\newcommand{\eeast}{\end{eqnarray*}}

\newcommand{\ind}{\mathbbm{1}}
\newcommand{\E}{\mathbb{E}}
\newcommand{\Prx}{\mathbb{P}}

\begin{document}
\title{Sample-path large deviations for stochastic evolutions driven by the square of a Gaussian process.}
\author{Freddy Bouchet}
\email{Freddy.Bouchet @ ens-lyon.fr}
\affiliation{Laboratoire de Physique
ENS de Lyon and CNRS
46, alley d'Italie
F-69364 Lyon cedex 07
France}
\author{Roger Tribe}
\email{r.p.tribe@warwick.ac.uk}
\affiliation{Department of Mathematics, University of Warwick, Coventry CV4 7AL, United Kingdom}
\author{Oleg Zaboronski}
\email{o.v.zaboronski@warwick.ac.uk}
\affiliation{Department of Mathematics, University of Warwick, Coventry CV4 7AL, United Kingdom}
\date{\today}
\begin{abstract}
Recently, a number of physical models has emerged described by 
a random process with increments given by a quadratic form 
of a fast Gaussian process. We find that the rate function which describes sample-path large deviations for such a process can be computed from the large domain size asymptotic of a certain Fredholm determinant. The latter can be evaluated 
analytically using a theorem of Widom which generalizes the celebrated Szeg\H{o}-Kac formula to the multi-dimensional case. This provides a large class of 
random dynamical systems with time scale separation for which
an explicit sample-path large deviation functional can be found. Inspired by problems in hydrodynamics and atmosphere dynamics, 
we construct a simple example with a 
single slow degree of freedom driven by
the square of a fast multi-variate  Gaussian process and analyse its large
deviations functional using our general results. Even though the noiseless limit
of this example has a single fixed point, the corresponding large deviations
effective potential has multiple fixed points. In other words, it is the addition
of  noise that leads to
metastability.  We use the explicit answers for the rate function
to construct  instanton trajectories connecting the metastable states.
\end{abstract}
\maketitle
\section{Introduction}

Large deviation theory recently became a key theoretical tool for the statistical mechanics of non equilibrium systems. Describing sample-path large deviations for the dynamics of effective degrees of freedom leads to a precise understanding of typical and rare trajectories of physical, biological or economic processes. A paradigm example for the effective descriptions of complex systems using large deviation theory is the macroscopic fluctuation theory of systems of interacting particles~\cite{bertini2015macroscopic}. However, for genuine non-equilibrium processes, without local detailed balance, the class of systems for which the rate function can be found explicitly is extremely limited. 

In this paper, we consider a class of systems for which the effective dynamics has increments which are given by a quadratic form of a fast Gaussian process. This  type of stochastic driving is relevant for many applications. Quadratic interactions are common in many physical examples such as hydrodynamics, plasmas described by the Vlasov equation, magneto hydrodynamics, self gravitating systems, the KPZ equation, quadratic networks (for instance heat transfer across quadratic networks~\cite{saito2011generating}), to cite just a few. For all these systems with quadratic nonlinearities, in some regime a separation of time scale exists and the effective degrees of freedom are coupled to fast evolving Gaussian processes. This is the case, for example, for the kinetic theories of plasma~\cite{Lifshitz_Pitaevskii_1981_Physical_Kinetics,Nicholson_1991}, self gravitating systems~\cite{Binney_Tremaine_1987_Galactic_Dynamics}, geostrophic turbulence~\cite{Bouchet_Nardini_Tangarife_2013_Kinetic_JStatPhys}, wave turbulence~\cite{nazarenko2011wave} for some specific dispersion relations, among many other examples. From a theoretical and mathematical perspective, modelling the driver of the effective degrees of freedom by a quadratic form of a fast Gaussian process proves to be  a decisive simplification. With this assumption, we will be able to write explicit formulae for the sample-path large deviation rate function, and proceed to its analysis in many interesting examples. 

The study of a slow process coupled to a fast one is a classical paradigm of physics and mathematics,
the celebrated Kapitza pendulum \cite{landau1976mechanics} being a canonical example. 
 For such fast/slow dynamics, one can study the averaging of the effect of the fast variable on the slow one (law of large numbers), or the typical fluctuations (stochastic averaging~\cite{pavliotis2008multiscale}), or the rare fluctuations described by the large deviations theory~\cite{freidlin2012random}. 
The latter is a natural tool for describing the evolution
of metastable systems consisting of long periods spent near an equilibrium point interspersed 
by rare transitions to a distinct equilibrium along an almost deterministic 'instanton' trajectory.
A number of systems with time-scale separation and drift
quadratic in fast variables exhibit metastability, see e. g.
\cite{bouchet}, \cite{Bouchet_Rolland_Simonnet_2019} and references therein.

The large deviation theory has been developed for slow/fast Markov processes ~\cite{freidlin1978averaging,veretennikov2000large} or deterministic systems~\cite{kifer1992averaging,kifer2004averaging}. Unfortunately, there are not many examples of fast/slow
systems for which the large
deviations rate function is known explicitly, which would enable the study of detailed properties of the systems such as the equilibrium points and transition
trajectories connecting them.

A pedagogical review of the large deviation theory for systems of stochastic differential equations (SDE's) with two well separated time scales can be found in \cite{thomas}. The theory is illustrated by a class of examples such that 
the drift for the slow process is given by a second degree polynomial of the fast process. The
corresponding large deviations principle is expressed in terms of the solution 
of a matrix Ricatti equation.
Unfortunately, the resulting expression is not explicit enough to study
the practically important phenomenon of {\em stochastically}
generated metastability: all metastable models considered in \cite{thomas} 
already possess multiple
fixed points in the noiseless limit. The turbulent models discussed in \cite{Bouchet_Laurie_Zaboronski_2014} suffer from the same flaw: metastability
appears due to a careful choice of the potential rather than being generated dynamically.  

The main new contribution of the current work is two-fold. Firstly, we apply the asymptotic theory of Fredholm determinants to the calculation of the large deviation rate function;
this results is an explicit formula for the rate function which characterizes sample-path large deviations for the slow process
in terms of a finite-dimensional determinant of a matrix of the size equal to the number of fast degrees of freedom. 
Essentially, Szeg\H{o}'s theory of Fredholm determinants is used to build
an asymptotic solution to the matrix Ricatti equation of \cite{thomas}. 

Secondly, we introduce a concrete illustrative example
with stochastically generated metastability. This is a
system of stochastic differential
equations with a single slow variable and a multi-dimensional fast variable for which all the drifts are quadratic such
that in the noiseless limit there is a unique fixed point. However, the addition of noise leads to the appearance of
multiple fixed points for the effective Hamiltonian dynamics describing the sample-path deviations. In other words the noisy system
exhibits metastability. We use our explicit knowledge of the rate function to construct transition paths (instanton
trajectories) between the fixed points.

A third result of our paper is of a more technical nature: as it turns out, it is enough to characterise
the fast variables as a Gaussian process with the auto-correlation function which decays sufficiently
fast, for example exponentially. In particular, it is not necessary to require that the fast
process be Markov. Relaxing this assumption opens up a possibility of using our results in turbulence modelling in the following way: the auto-correlation
functions of the small scale turbulence are measured experimentally and
used to model the small scale fluctuations as a fast Gaussian process.
Then the large deviations properties
of the large scale turbulence can be studied theoretically using the theory described below. 
A rigorous validation of the large deviations principle without assuming Markovianity is also a very natural   question for the probability theory. 

It is worth stressing that the current paper does not deal with applications of the developed theory to specific
physical systems. However, it has been already proved useful for the study of large deviations in a mean field
model of plasma kinetics, see the recent preprint \cite{feliachi2021dynamical} for details. We also hope that
we can apply the  
explicit formulas found here to understand the hydrodynamic bistability
discussed in 
~\cite{bouchet,Bouchet_Rolland_Simonnet_2019}. 

The rest of the paper is organised as follows. We start with the definition of the model in Section \ref{secmodel} and give a heuristic
derivation of the corresponding large deviation principle in Section \ref{ldp}. The highlight
of this section is the application of Widom's theorem for the asymptotics of Fredholm determinants
to the calculation of the rate function. In Section  \ref{example} we show the emergence
of metastability for a particular representative of our class of models and study the corresponding
'instanton' trajectories. Brief conclusions are presented in Section \ref{concl}.
Appendices \ref{app1}, \ref{app2}, \ref{appLB}  contains some technical derivations for Section \ref{ldp}.
Appendix \ref{app3} contains a review of Widom's theorem.

\section{Slow dynamics quadratically driven by a fast Gaussian process}\label{secmodel}
Consider the following stochastic model:
\bea\label{modeleq}
\left\{
\begin{array}{ccl}
\dot{X}(t)&=&Y^T\left(\frac{t}{\epsilon},X(t)\right)MY\left(\frac{t}{\epsilon},X(t)\right)-\nu X(t),\\
X(0)&=&x_0,
\end{array}
\right.
\eea
where $\{X(t)\}_{t\geq 0}$ is an $\R^n$-valued random process,
$\epsilon$ is a parameter, which determines time scale separation between
the processes $X$ and $Y$,  $0<\epsilon<<1$; for a fixed $x \in \R^n$, $(Y(t,x), t,x \in \R)$
is an $N$-dimensional time-stationary centred Gaussian process with the auto-correlation
function (covariance matrix)
\bea\label{covmod}
&C_{ij}(\tau,x,y)=\E\left[Y_i(t,x)Y_j(t+\tau,y) \right],\\
&\text{where }\tau \geq 0, 1\leq i,j \leq N,\nonumber
\eea 
which is assumed to be continuous in all the arguments $\tau, x, y$. 
As we will see, only $C(\tau,x,x)$ enters the final
expression for the large deviation rate function, which justifies our shorthand notation $C(\tau,x):=C(\tau,x,x)$.
Finally, $M$ is a $n\times N\times N$ matrix, symmetric with respect to the 
permutation of the last two indices, and $\nu>0$ is a parameter.  Notice that the $(X,Y)$ process need not be Markov.

We assume that $C(\tau, x)$ decays at least  exponentially with $\tau$, perhaps uniformly
with respect to $x$. 
Then, in the limit of $\epsilon\rightarrow 0$, the slow random process $X$ stays near
the solution to the deterministic equation
 \bea\label{ccl}
\left\{
\begin{array}{ccl}
\dot{x}(t)&=&\text{tr}\left[ M C(0,x(t))\right]-\nu x(t),\\
x(0)&=&x_0,
\end{array}
\right.
\eea
where $\text{tr}$ is the trace over $N$ 'fast' indices. Equation (\ref{ccl})
is a consequence of the ergodic average applied to the integral form of (\ref{modeleq}). 
The {\em typical} fluctuations of $X(t)$ around $x(t)$ are Gaussian, with
covariance of order $\epsilon$ (more precisely, the distribution of $\lim_{\epsilon\rightarrow 0}$
$\frac{X(t)-x(t)}{\sqrt{\epsilon}}$ is centred Gaussian). Here
we are interested in the statistics of {\em large deviations} of $X(t)$ when $X(t)-x(t)=O(1)$, which are no longer Gaussian in general.
\section{Large deviation principle for paths of the slow process}\label{ldp}
If the fast process were Markov, the starting point for our analysis would be the known large deviations
principle for fast-slow Markov systems expressed in terms of the Legendre transform  of the cumulant generating functional
\bea\label{cumgen}
Z_T[x,\lambda]\!\!=\!\!\log \E_{Y} \exp\left\{\!\!\int_0^T \!\!\!\!dt \lambda(t)f\left(x(t),Y(t,x(t)) \right) \right\},
\eea
where $f$ is the right hand side of the equation (\ref{modeleq}) 
for the slow degrees of freedom, see \cite{thomas} for a review.

However, it turns out that assuming the Gaussianity of $Y$ and the exponential decay of 
the corresponding auto-correlation function it 
is possible to arrive at a counterpart of (\ref{cumgen}) without assuming Markovianity, see 
Eq. (\ref{step0}) below. As we already explained in the introduction, by extending the range 
of possible drivers
we open up a possibility of applying our results to turbulent modelling. 

The following is essentially a computation 
of the functional integral measure for the slow variable $X$, 
which we have to use instead of
the Martin-Siggia-Rose method \cite{msr}, which is only applicable to the Markov case.  
It is not a  proof, but rather a heuristic argument devised to give an intuitive feel for
the conjectured form of the large deviation principle. 
Let us fix the final time $t>0$, choose
a large integer $P \in \N$ and a positive number $\eta$ and
define $$\Delta t=\frac{t}{P}, ~b_\eta(x)=\prod_{\alpha=1}^n[x_\alpha-\eta,x_\alpha+\eta],$$ 
where $x$ is a point in $\R^n$ and $\prod$ stands for the direct product of intervals. 
Geometrically, $b_\eta(x)$ is a hypercube in $\R^n$ centred on $x$
with  side $2\eta$.
Let $(\lambda_1, \lambda_2,\ldots, \lambda_P)$, $(x_1, x_2, \ldots, x_P)$
be two sequences of $n$-dimensional vectors. 
Let $\Prx$ be the probability distribution for the process $(X,Y)$.
Let $\E$ be the corresponding expectation.
We are interested in the probability that at the times $k\Delta t$
the corresponding values of the slow process $X(k\Delta t)$ are near the points $x_k$
, $1\leq k\leq P$. A computation exploiting Chebyshev's inequality shows that for any sequence of $\lambda$'s
\bea
&&\epsilon \log\Prx\left[X(k \Delta t)\in b_\eta(x_k), k = 1,\ldots, P \right]\leq
\label{step0}\\
&&\sum_{k=1}^P\!\!\bigg(\!\! \lambda_k^T\left(x_{k-1}\!-\!x_{k}\right)
\!\!+\!\!\epsilon \log \E
\left[ e^{\lambda_k^T F(Y,x_{k-1})}\right]\!\bigg)\!\!
+\!\!R(\epsilon,\Delta t,\eta),\nonumber
\eea
where 
\bea\label{forcing}
&&F(Y,x)=\int_{0}^{\Delta t/\epsilon}d\tau Y^T(\tau,x) MY(\tau,x)
-\nu x \Delta t/\epsilon,
\eea
and $R$ is an error term depending on $\epsilon$, $\eta$ and $\Delta t$ such that for
$\eta=\mu\Delta t$,
\bea\label{errest}
\lim_{\mu\rightarrow 0} \lim_{\Delta t\rightarrow 0}\lim_{\epsilon \rightarrow 0} R(\epsilon,\Delta t,\mu \Delta t)=0.
\eea
The derivation of (\ref{step0}) is based on the approximation of $Y$
by a bounded process with a finite dependency range. It is 
carried out in  Appendix \ref{app1}. Here we would only like to point out that
the dependence on $\lambda$ in the right hand side of (\ref{step0})
appears due to the repeated use of Chebyshev's inequality. Intuitively, the sequence $(\lambda_k)$ 
is the discretised counterpart of the response field appearing in Martin-Siggia-Rose computation.

The next aim is to compute the expectation  
$$
\E
\left[ e^{\lambda^T F(Y,x)}\right]\!\!=\!e^{-\frac{\Delta t}{\epsilon}\nu \lambda^Tx}
\E\left[ 
e^{\lambda^T
\left(\int_{0}^{\Delta t/\epsilon}d\tau Y^T(\tau,x) MY(\tau,x)
\right)}
\right],
$$
which can be done using the fact that for a fixed $x\in \R^n$ the process
$Y(\cdot,x)$ is stationary and Gaussian. What follows is the key computation of the paper
linking averaging over fast Gaussian fields with the asymptotic of certain Fredholm
determinants. 
Let us define $m:=\lambda^T M$, an $N\times N$
symmetric matrix. It can be  decomposed as  $m=S^TS$, where $S$ is a possibly complex
Cholesky factor of $m$.
 Rewrite  
\beast
&&\exp\left[\lambda^T
\left(\int_{0}^{\Delta t/\epsilon}d\tau Y^T(\tau,x) MY(\tau,x)
\right)\right]
\\
&&=\int \prod_{\tau}\frak{D}q(\tau) e^{-\frac{1}{4}\int_0^T d\tau
q^T(\tau)q(\tau)+\int_0^{\Delta t/\epsilon}d\tau q^T(\tau) SY(\tau,x)}
\eeast
(Hubbard-Stratonovich transformation). Then, for sufficiently small components of $\lambda$,
\bea
&&\E \exp \left[ \lambda^T
\left(\int_{0}^{\Delta t/\epsilon}d\tau Y^T(\tau,x) MY(\tau,x)
\right) \right]
\nonumber\\
&&=\!\!
\int \!\! \prod_{\tau} \!\!\frak{D}q(\tau) e^{-\frac{1}{4}\int_0^{\frac{\Delta t}{\epsilon}} \!\!\!\!d\tau
q^T(\tau)q(\tau)}\E\left(e^{\int_0^{\frac{\Delta t}{\epsilon}}\!\!\!\!d\tau q^T(\tau) SY(\tau,x)}\right)
\nonumber\\
&&=\!\!\int  \prod_{\tau} \frak{D}q(\tau) \exp \bigg[-\frac{1}{4}\int_0^{\Delta t/\epsilon} d\tau
q^T(\tau)q(\tau)
\nonumber\\
&&+\frac{1}{2}
\int_0^{\Delta t/\epsilon} \!\!\!\!d\tau_1\int_0^{\Delta t/\epsilon}\!\!\!\!d\tau_2 
q^T(\tau_1)SC(\tau_1-\tau_2,x)S^Tq(\tau_2) \bigg]
\label{fdet1}\\
&&=
\text{Det}^{-\frac{1}{2}}\!\!\left(I\!-\!2S\hat{C}_{\Delta t/\epsilon}(x)S^T\right)
=
\text{Det}^{-\frac{1}{2}}\!\!\left(I\!-\!2 m\hat{C}_{\Delta t/\epsilon}(x)\right).
\nonumber
\eea
Here $m\hat{C}_{\Delta t/\epsilon}(x)$ is an integral operator acting on (square integrable) $\R^N$-valued
functions on $[0,\Delta/\epsilon]$ as follows:
\bea
&&f_\alpha(t)\!\!\mapsto \!\! m\hat{C}_{\Delta t/\epsilon}(x)(f)_\alpha(t)\!\!
\nonumber\\
&=&\!\!\sum_{\beta,\delta=1}^N\int_0^{\Delta t/\epsilon} d\tau m_{\alpha \beta}C_{\beta, \delta}(t-\tau,x)f_{\delta}(\tau),
\eea
for all $\alpha=1,\ldots,N; t \in [0,\Delta t/\epsilon]$.

In what follows we will use capital $\text{Det}$ and $\text{Tr}$ to denote operator determinant
and trace, and reserve the lowercase $\det$ and $\text{tr}$ for the determinant and the trace of finite-dimensional matrices.

The calculation of (\ref{fdet1}) in the limit $\epsilon\rightarrow 0$  
requires the asymptotic analysis of the Fredholm 
determinant of an integral operator acting on functions defined on a large interval.
Fortunately, such an asymptotic can be computed using Widom's theorem,
which generalises the celebrated Szeg\H{o}-Kac  formula for Fredholm determinants,
see \cite{widom}: for a sufficiently small $m$ (e.g. with respect to a matrix norm),
\bea
&&\log \text{Det}\left(I-2 m\hat{C}_{\Delta t/\epsilon}(x)\right)
\nonumber\\
&=&\frac{\Delta t}{\epsilon}\int_\R \frac{dk}{2\pi}\log 
\det \left(I-2 m\tilde{C}(k,x)\right)+O\left(1 \right),
\label{fdet2}
\eea
where 
$
\tilde{C}(k,x)=\int_\R d\tau e^{ik\tau} C(\tau,x)
$
is the Fourier transform of the autocorrelation function $C(\tau,x)$. This remarkable statement
is reviewed in Appendix \ref{app3}.
Substituting (\ref{fdet1}), (\ref{fdet2}) into (\ref{step0}) we find
\begin{widetext}
\begin{eqnarray}
&\epsilon&\!\!\!\! \log\Prx\left[X(p \Delta t)\in b_\eta(x_p), p = 1,\ldots, P \right]
\nonumber\\
&\stackrel{(\ref{step0})}{\leq}& 
\sum_{p=1}^P  \Delta t
\lambda_p^T\left(\frac{x_{p-1}-x_{p}}{\Delta t}-\nu x_{p-1} \right)
+\sum_{p=1}^P\epsilon \log \E \exp\left[\int_{0}^{\Delta t/\epsilon}d\tau 
Y^T(\tau,x_{p-1})mY(\tau,x_{p-1}) \right]+R
\nonumber
\\
&\stackrel{(\ref{fdet1})}{=}&
\sum_{p=1}^P  \Delta t
\lambda_p^T\left(\frac{x_{p-1}-x_{p}}{\Delta t}-\nu x_{p-1} \right)
-\frac{1}{2}\sum_{p=1}^P \epsilon \log \text{Det}\left(I-2m\hat{C}_{\frac{\Delta t}{\epsilon}}(x_{p-1})\right)
+R
\nonumber\\
&\stackrel{(\ref{fdet2})}{=}&
\sum_{p=1}^P  \Delta t
\lambda_p^T\left(\frac{x_{p-1}-x_{p}}{\Delta t}-\nu x_{p-1} \right)
-\frac{1}{2}\sum_{p=1}^P \epsilon \left[\frac{\Delta t}{\epsilon}
\int_\R\frac{dk}{2\pi}\log 
\det\left(I-2m\tilde{C}(k,x_{p-1})\right)
\right]
+R+O(\epsilon P)
\nonumber\\
&=&
\sum_{p=1}^P  \Delta t
\lambda_p^T\left(\frac{x_{p-1}-x_{p}}{\Delta t}-\nu x_{p-1} \right)
-\frac{1}{2}\sum_{p=1}^P \Delta t \int_\R\frac{dk}{2\pi}\log 
\det\left(I-2\lambda_p^TM\tilde{C}(k,x_{p-1})\right)
 +R+O(\epsilon P),
\label{step1}
\end{eqnarray}
\end{widetext}
where the $O(\epsilon P)$ addition to the error term comes from the $O(1)$ term in (\ref{fdet2}).
The expression (\ref{step1}) is an upper bound on the (discretisation of) the functional integral measure
for the process $X$. 


The next step is akin to the calculation of a path integral for $\epsilon \rightarrow 0$ using 
the Laplace method.
The question we ask is: what is the probability $\Prx[X \in D]$, where $D$ is a `nice' subset 
of the space $C([0,t],\R^n)$ of 
$\R^n$-valued functions on $[0,t]$?

By analogy with the finite-dimensional Laplace method, one needs to minimize
the functional integral measure (\ref{step1}) over $D$. The details of this computation
can be found in the Appendix \ref{app2}. Here we will just state the  answer after
taking the continuous limit $\Delta t\rightarrow 0$: 
\bea \label{step12}
&&\limsup_{\epsilon \rightarrow 0} \epsilon \Prx[X \in D]\leq
\!\!-\!\!\inf_{x \in D}\bigg[
\int_0^t  d\tau
\lambda_p^T(\tau)\left(\dot{x}(\tau)\!+\!\nu x(\tau) \right)
\nonumber\\
&&+\!\frac{1}{2} \int_\R\frac{dk}{2\pi}\log 
\det\left(I\!-\!2\lambda^T(\tau)M\tilde{C}(k,x(\tau))\right)\bigg].
\eea
The derived bound is valid for an arbitrary function $\lambda$. Taking the infimum
of the right hand side of (\ref{step12}) over this function, one  gets the
optimal upper bound on $\Prx[X \in D]$:
\bea\label{step13}
&&\limsup_{\epsilon \rightarrow 0} \epsilon \Prx[X\!\! \in \!\!D]\!\!\leq
\!\!-\!\!\sup_\lambda\inf_{x \in D}\!\!\bigg[\!\!
\int_0^t  \!\!d\tau
\lambda_p^T(\tau)\big(\dot{x}(\tau)\!
\nonumber\\
&&+\!\nu x(\tau) \big)
\!\!\!+\!\frac{1}{2} \!\!\int_\R\!\!\frac{dk}{2\pi}\log \!
\det\!\!\left(I\!-\!2\lambda^T\!\!(\tau)M\tilde{C}(k,x(\tau))\right)\!\!\bigg].
\eea
Staying at the similar level of rigour and
using the same set of assumptions about the fast process as above, one 
can show that the r. h. s. of  (\ref{step13}) is also a lower
bound on $\liminf_{\epsilon \rightarrow 0} \epsilon \Prx[X\!\! \in \!\!D]$.
The corresponding calculation is based on a standard trick of deforming the probability distribution
in such a way that the low probability event at hand becomes almost inevitable, see e. g. 
\cite{durrett} for a short introduction. The details are given in Appendix \ref{appLB}.

Therefore it is natural to conjecture that the slow
process $X$ satisfies the large deviation principle with rate $\epsilon$ and
the explicit rate function given by 
\bea
&&S_{eff}[\lambda,x]=
\int_0^td\tau \lambda^T(\tau)\left(\dot{x}(\tau)+\nu x(\tau)\right) 
\nonumber\\
&+&\frac{1}{2}\!\! \int_0^t \!\!\!\!d\tau\!\!
\int_\R \!\!\frac{dk}{2\pi} \log \det \left(I\!-\!2\lambda^T(\tau) M\tilde{C}(k,x(\tau))\right),
\label{seff}
\eea
provided that $t$ is not too large.
Less formally one can write
\bea
\Prx\left[X \in D \right]\sim e^{-\frac{1}{\epsilon}\sup_\lambda\inf_{x\in D}S_{eff}\left[\lambda,x\right]}
\eea
A typical application of the rate functional guessed above is the estimation 
of the probability of
transitioning between fixed points of the typical evolution (\ref{ccl}). If $x_0, x_1$ are two
such points, then
\bea
\Prx\left[X(t)\in dx_1\!\!\mid\!\! X(0)=x_0 \right]
\!\!\sim \!\!
e^{-\frac{1}{\epsilon}
\sup_\lambda\inf_{x}S_{eff}[\lambda,x]},
\label{ldm}
\eea
where the $\inf$ and the $\sup$ are taken over the functions $x,\lambda$  on $[0,t]$
such that $x(0)=x_0, x(t)=x_1$.

When analysing specific examples, it is often convenient to think of (\ref{seff}) as the 
action functional for a mechanical system with generalized coordinates $x$ and generalized momenta $\lambda=\frac{\delta S_{eff}}{\delta \dot{x}}$.
This system is Hamilton's with the Hamiltonian
\bea\label{heff}
\!\!\!\!\!\!\!H_{eff}(\lambda,x)\!\!=\!\!-\nu\lambda^Tx\!\!-\!\!\!
\int_\R \frac{dk}{4\pi} \log \det\!\! \left(I\!\!-\!\!2\lambda^T \!\!M\tilde{C}(k,x)\right)\!\!,
\eea
see \cite{landau1976mechanics} for details of the map between the Lagrangian and the Hamiltonian formalisms. 
As a self-consistency check, let us verify that the average evolution equation (\ref{ccl})
appears as an equation for a typical trajectory for the large deviation principle
(\ref{ldm}), (\ref{seff}). A typical trajectory $(\lambda_c,x_c)_{0\leq \tau \leq t}$
is a solution to Euler-Lagrange equations associated with $S_{eff}$ such that
\beast
S_{eff}[\lambda_c,x_c]=0.
\eeast
Examining the derivation of the large deviation principle, it is reasonable to expect
 that $\lambda_c=0$. Expanding (\ref{seff}) around $\lambda=0$ we find
 \beast
 S_{eff}\!\!=\!\!\int_0^t\!\!\!\!d\tau \lambda^T(\tau)\bigg(\!\!\dot{x}(\tau)
 \!\!+\!\!\nu x(\tau)\!\!-\!\!\text{Tr}\left[MC(0,x(t)) \right]\!\! \bigg)\!\!+\!\!O(\lambda^2),
 \eeast
 where we used that $\int_\R \frac{dk}{2\pi} \tilde{C}(k,x)=C(0,x)$. Therefore,
 $\lambda=0$ solves the Euler-Lagrange equations if 
 \beast
 \dot{x}(\tau)
 +\nu x(\tau)-\text{Tr}\left[MC(0,x(t)) \right]=0,~x(0)=x_0,
 \eeast
 which coincides with (\ref{ccl}).
In particular, the fixed points of the slow dynamics are solutions to 
 \bea\label{sfp}
 \nu x=\text{Tr}\left[MC(0,x) \right]
 \eea
{\bf Remarks.}
\begin{enumerate}
\item If $N=1$, and $Y$ solves an Ornstein-Uhlenbeck SDE with $X$-dependent drift, the corresponding
large deviation principle was derived in \cite{bouchet} and is consistent with conjecture (\ref{step13})
for all values of $\lambda$. However,
in general one has to check that the optimal $\lambda$ belongs to the domain of applicability
of Widom's theorem, which is one of the challenges for the rigorous justification of the conjecture. A natural
guess is that the  minimizer must be small enough to ensure positive definiteness of the quadratic form in the functional integral
(\ref{fdet1}). 
\item If $Y$ appears as a solution to an Ornstein-Uhlenbeck system of stochastic
differential equations,  then (\ref{step13}) can be viewed as a solution 
to the matrix Riccatti problem for the rate function derived in \cite{thomas}.
\item In the context of modelling of two-dimensional turbulent flows, equation
(\ref{modeleq}) can be interpreted as follows: $Y$ is a Gaussian model of fast small-scale
velocity field whose evolution depends on the static background created by $X$; 
$X$ is a large scale velocity field slowly evolving under the influence of $Y$. Thus the model
can be thought of as a non-linear generalisation of the passive vector advection model. 
The shape of $C$ reflects the nature of the small scale turbulent flow (compressibility, isotropy, etc.)
\end{enumerate}
\section{An example inspired by multistability in hydrodynamic and geostrophic turbulence}\label{example}

The aim of this section is to present an example of the use of the large deviation principle (\ref{seff}). We are specifically interested in metastability phenomena observed in two dimensional~\cite{bouchet} and geostrophic~\cite{Bouchet_Rolland_Simonnet_2019} turbulent flows. In previous works, we have studied metastability for geostrophic dynamics~\cite{Bouchet_Laurie_Zaboronski_2014}, in cases when the turbulent flows is forced by white noises, and the stochastic process is an equilibrium one with detailed balance or generalized detailed balance. The large deviation principle (\ref{seff}) opens the possibility for studying metastability for turbulent flows modelled as a non-equilibrium process. 
As a first step, we will now demonstrate the stochastic generation of metastability
for systems with time scale separation  
using the simplest example of the system of stochastic differential equations with  
the quadratic drift for the slow variable. 

To formulate the example, it will be easier to use complex notations. The fast variable
$Y(\cdot, x) \in \C^N$ will be an analogue of the set of Fourier components that describe the turbulent fluctuations. $Y$ is the stationary solution of the complex Ornstein-Uhlenbeck process.
The SDE for the full fast-slow system is as follows:
\bea\label{sys}
\left\{
\begin{array}{l}
dY(t,x)=-\Gamma(x) Y(t,x)dt+\sigma dW(t),\\
\!dX\!(t)\!=\!Y\!(t/\epsilon,X\!(t))^*MY\!(t/\epsilon,X\!(t))dt\!-\!\nu X\!(t)dt,
\end{array}
\right.
\eea
where $M$ is an $n\times N\times N$ matrix self-adjoint with respect to the last two indices;
$dW$ is the $\C^N$-valued Brownian motion,
with the non-trivial covariance
\bea
d\overline{W}_i dW_j=\delta_{ij}dt,
\eea
$\Gamma(x)$ is a complex matrix, whose  eigenvalues have positive real parts,
\bea
\Gamma(x)=\Gamma^{(0)}+ix^T \Gamma^{(1)},
\eea
where $\Gamma^{(0)}$ is a real positive definite $N\times N$ matrix, $\Gamma^{(1)}$
is a real $n\times N\times N$ matrix. The former describes dissipation, whereas the
latter corresponds to the `rotational' advection of $Y$ by the slow field $X$.
All the coefficients are polynomials of degree at most one in $x$. 

The model (\ref{sys}) is a representative of the class of models (\ref{modeleq}), (\ref{covmod})
treated in this paper: the slow field  $X$ is driven by a quadratic form of 
$Y(\cdot,x)$ which, as follows from the first of the SDE's (\ref{sys}), is  
Gaussian with the exponentially decaying autocorrelation  function. In particular,
the large deviations rate function can be derived from Eq. (\ref{seff}) of the previous section. Finally,
notice that the process $(X,Y)$ defined by (\ref{sys}) is Markov. As pointed out at the end of Section \ref{ldp},
the task of justifying the large deviations principle (\ref{seff}) reduces in this case just to the check of the applicability
of Widom's theorem, whereas the important intermediate result (\ref{step0}) can be established rigorously, see \cite{veretennikov2000large}.

The structure of 
the system of SDE's (\ref{sys}) resembles that of the quasi-linear approximation
to the Navier-Stokes equation or quasigeostrophic equations, 
see \cite{Bouchet_Nardini_Tangarife_2013_Kinetic_JStatPhys} for details: 
the non-linearity in the right hand side 
is quadratic, the evolution of the slow variable is driven by the term quadratic
in the fast variable, the drift of the fast variable resembles  advection
by the slow field $X$. Let us stress that the model does not have any artificial ``built-in'' non-linearity:
the noiseless limit of (\ref{sys}) has a unique critical point $X=Y=0$. The metastability
described below is a purely stochastic  effect.

Some standard computations
lead to formulae for the correlation and auto-correlation functions, 
$C(0,x):=\E(Y(0,x)\otimes Y^*(0,x))$, $C(\tau,x):=\E(Y(\tau,x)\otimes Y^*(0,x))$. Here
$\otimes$ denotes tensor product: for vectors $a,b \in \C^n$, $a\otimes b$ is an $n\times n$
matrix such that $(a\otimes b)_{ij}=a_ib_j$.
$C(0,x)$ solves the Lyapunov equation,
\bea
\Gamma(x)C(0,x)+C(0,x)\Gamma^*(x)=\sigma \sigma^*,
\eea
whereas
\bea
C(\tau,x)=e^{-\Gamma(x) \tau} C(0,x),~\tau\geq 0.
\eea
If $\tau<0$, then
$C(\tau,x)=C(0,x)e^{\Gamma^*(x) \tau}$.
The effective Hamiltonian (\ref{heff}) re-written in complex terms is
\bea\label{ham}
\!\!\!H_{eff}(\lambda, x)\!\!=\!\!-\lambda^T \nu x\!\!-\!\!\!\!\int_\R\frac{dk}{4\pi} 
\log \det\!\!
\left(I\!\!-\!\!2\lambda^TM\tilde{C}(k,x)  \right)\!\!,~~
\eea 
where
\bea
&&\tilde{C}(k,x):=\int_\R d\tau e^{ik\tau} C(\tau)
\nonumber\\
&&=
(\Gamma(x)-ik)^{-1}C(0,x)\!\!+\!\!C(0,x)(\Gamma^*(x)+ik)^{-1}\!\!,
\eea
is the Fourier transform of the auto-correlation function. 

Keeping matters as simple as possible, let us choose
$\Gamma^{(0)}$ and $\Gamma^{(1)}$ to be the diagonal matrices
with real entries $\{\gamma^{(0)}_p, \gamma^{(1)}_p\}_{1\leq p \leq N }$,
where $\gamma^{(0)}$'s are all positive. 
The fixed point equation (\ref{sfp}) takes the form 
\bea
&\sum_{j,k=1}^N \frac{\left(\sigma \sigma^*\right)_{jk} (M_\alpha)_{kj}}
{\left(\gamma^{(0)}_j+\gamma^{(0)}_k+i\sum_{\beta=1}^n \left[(\gamma_\beta^{(1)})_j-
(\gamma_\beta^{(1)})_{k}\right]x_{\beta} \right)}
\nonumber\\
&=\nu x_\alpha~, 1\leq \alpha \leq n.
\label{fpee}
\eea
Notice that if either the noise covariance matrix $\sigma\sigma^*$, or the interaction
matrix $M_\alpha$ is diagonal, there is a unique solution for the $\alpha$-th component of
the fixed point. Indeed, if $(M_{\alpha})_{kj}=0$ for  all $k\neq j$, then the left hand side
of equation (\ref{fpee}) becomes $x$-independent and the equation becomes linear w.r.t $x_\alpha$. The same
remark applies if $\sigma\sigma^*$ is diagonal. Similarly, the fixed point is unique if $(\gamma_\beta^{(1)})_j-
(\gamma_\beta^{(1)})_{k}=0$ for all $j,k, \beta$.  However, for general correlated noise, interaction
and an inhomogeneous rotation matrix $\gamma^{(1)}$, there are typically multiple solutions
to (26). 

\begin{figure}
\centering
\includegraphics[trim={3.8cm 0 0 0},clip, scale=0.2]{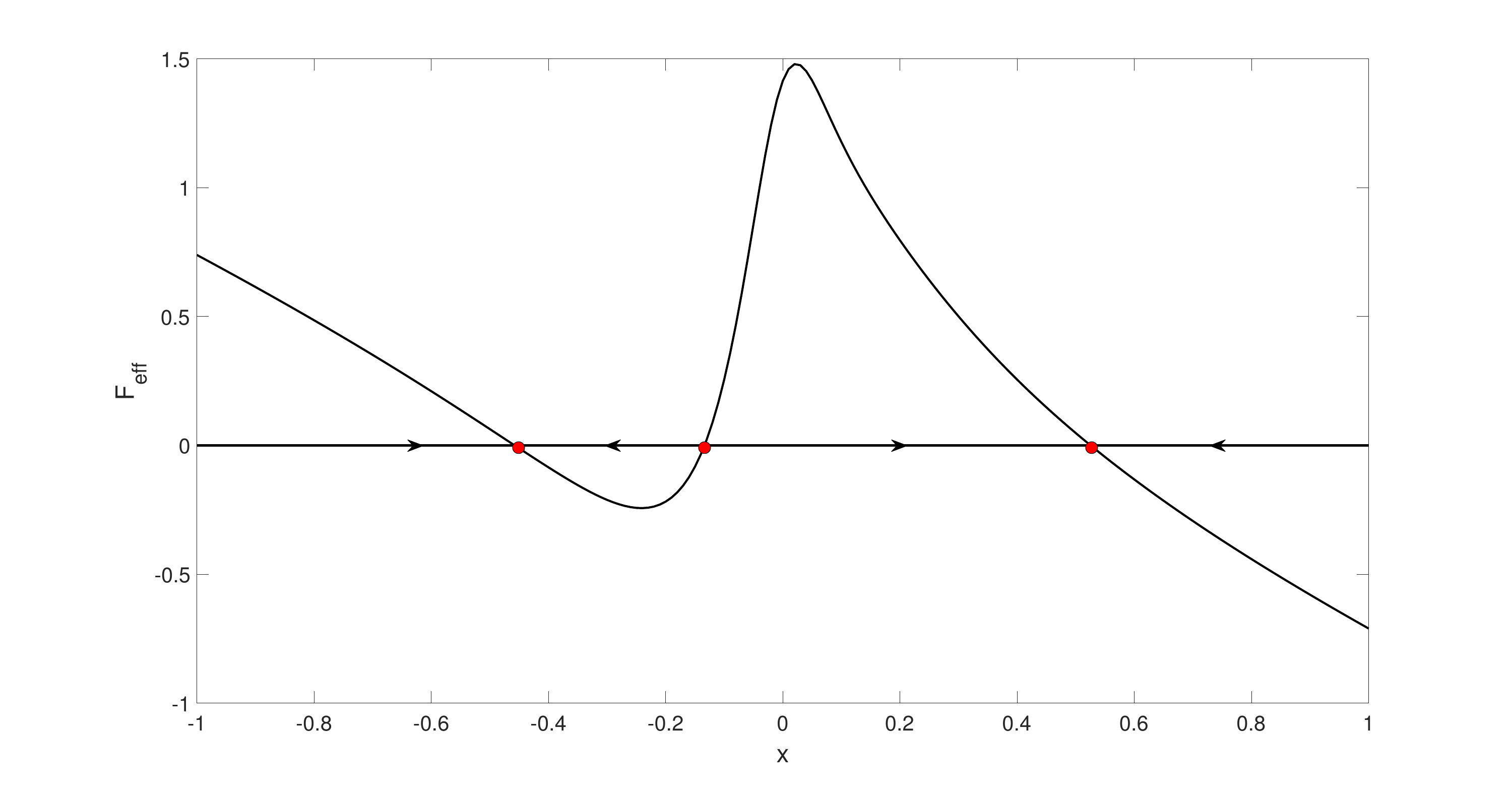}
\caption{The effective force $F_{eff}(x):=\frac{\partial H_{eff}}{\partial \lambda}(0,x)$ for the model (\ref{model}). Notice a pair of stable
fixed points of the averaged dynamics separated by an unstable fixed point.}
\label{fig1}
\end{figure}

We therefore conclude with the simplest non-trivial example 
such that (\ref{fpee}) has multiple real solutions, see Fig. \ref{fig1}.  For this example, $n=1$, $N=3$, $\nu=I_{3\time 3}$
and it has two stable and one unstable fixed points.
\bea
&&\sigma=\frac{1}{2^{1/4}}\left(\begin{array}{ccc}
-1-i&0 &0\\
1-i&-1-i &0\\
-1-i&1-i &-1-i
\end{array}
\right)\!\!,
\nonumber\\
&&M=\left(\begin{array}{ccc}
0&1 &1\\
1&0 &1\\
1&1 &0
\end{array}
\right) \!\!,
\gamma^{(0)}=
\left(\begin{array}{c}
1\\
1\\
1
\end{array}
\right)\!\!,
\gamma^{(1)}=\pi^2
\left(\begin{array}{c}
1\\
2\\
3
\end{array}
\right)\!\!.
\label{model}
\eea
The appearance of powers of $2$ and $\pi$ in the above parameterisation has no
special meaning. The choice $M_{ii}=0$ and $M_{ij}=const$ for $i\neq j$
reflects some properties of the interaction matrix for the $2$-dimensional Navier-Stokes
equation, but it is also not essential for the appearance of multiple equilibria.

\begin{figure}
\centering
\includegraphics[trim={1.8cm 0 0 0},clip,scale=0.33]{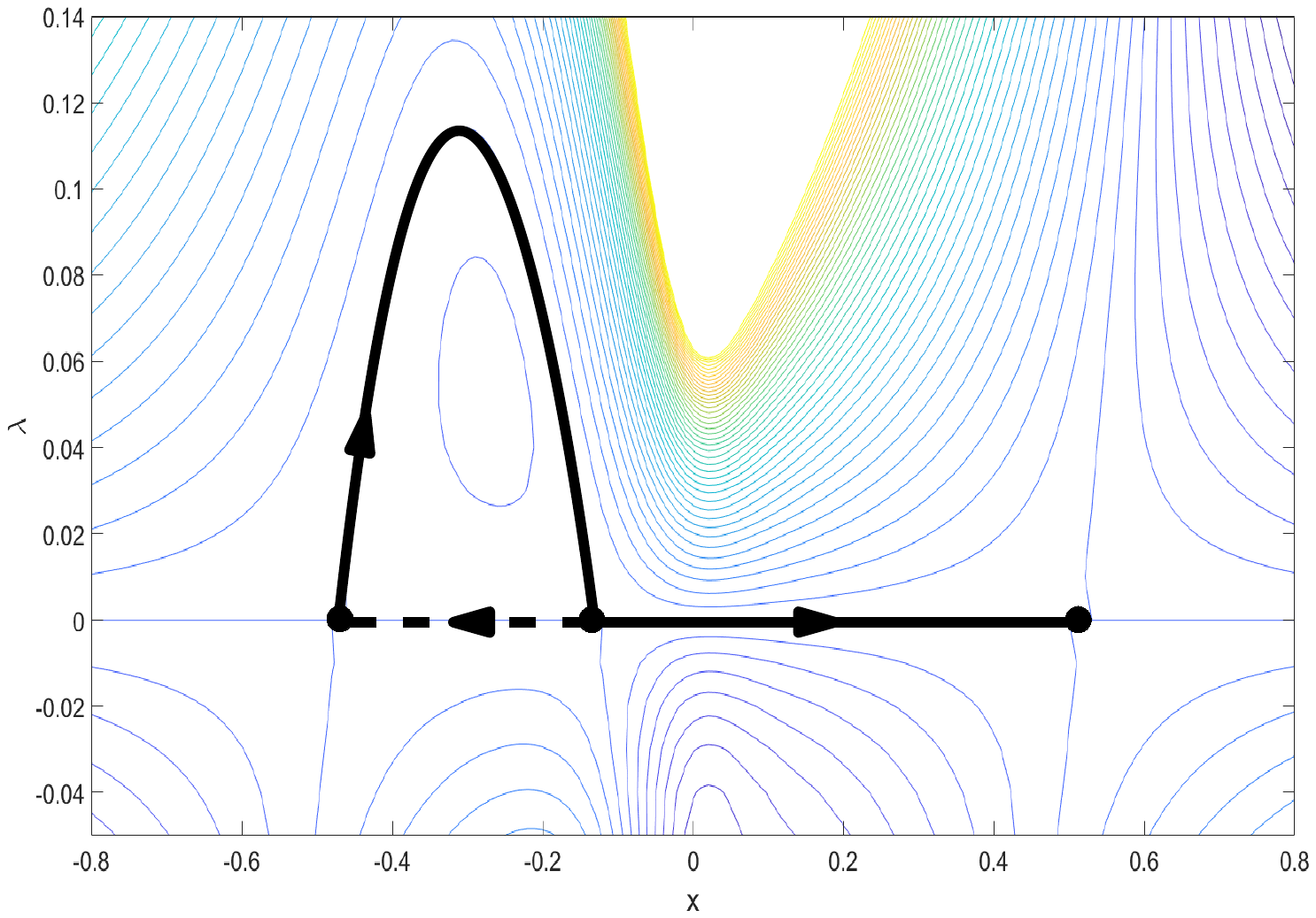}
\caption{Contour lines of $H_{eff}$ for the model (\ref{model}). The contour lines
in the upper half plane serve as optimal trajectories for transitions between the stable
fixed points in a finite time. The wide curve is the infinite-time optimal transition 
curve. The dashed segment marks the typical trajectory connecting the unstable and stable
fixed points.}
\label{fig2}
\end{figure}

The fact that multiple equilibria appear naturally in the
model (\ref{sys}) together with its link to quasi-linear hydrodynamics explained above 
makes us hope that the large deviation principle (\ref{seff}) might prove useful in
studying realistic hydrodynamic phenomena of metastability, such as the
zonal-dipole transition discovered in \cite{bouchet}. 
Euler-Lagrange equations associated with the effective action functional (\ref{seff})
are Hamilton's with the Hamiltonian (\ref{ham}). Therefore, each solution
lies on a constant energy surface $H_{eff}(\lambda,x)=E$. If there is a single slow variable,
the trajectories coincide with constant energy surfaces. This allows one to determine a family
of the most
likely transition paths between the fixed points (the instanton trajectories) by building
the contour plot of $H_{eff}$ numerically, see Fig. \ref{fig2}.
\section{Conclusions. Outlook}\label{concl}
Motivated by hydrodynamic applications, we have considered a model with two-time scales,
where the slow variable is driven by a quadratic function of a fast Gaussian process with
rapidly decaying auto-correlations. A natural question of
computing the probabilities of rare events in  this model reduces to the computation
of large-interval asymptotics for a certain Fredholm determinant. To the leading order, 
such a computation can be easily carried out using Widom's theorem. To apply the resulting
large deviation principle, we considered a special case of the fast field being a
complex Ornstein-Ohlenbeck process with the the rotational component of the drift given
by a linear function of the slow process. As it turns out, the average slow dynamics for such a model exhibits multiple equilibria, the transitions between which can be studied using large deviation theory. 

There are many natural further questions to ask. Firstly, it should be a straightforward
task to furnish a rigorous proof or provide a counter-example to the statement of the conjecture
(\ref{seff}). Secondly, for the cases, when the fast process conditional on the value of the slow process is
an Ornstein-Uhlenbeck process, it might be interesting to consider finite-$\epsilon$
corrections to the leading order answer. Albeit known, the sub-leading terms in the
Widom asymptotic are only characterised as solutions to  a certain matrix Wiener-Hopf 
integral equation. There is however a chance of finding these corrections rather 
more explicitly as solutions to time-dependent Riccatti equations derived in \cite{thomas}.

Finally, the model considered has the general structure of many equations of hydrodynamics, plasma dynamics, self-gravitating systems, wave turbulence or other physical system with quadratic couplings or interactions. It would therefore be extremely interesting
to analyse metastability for such physical systems, in the presence of time scale separation, using the findings of the present paper. 
\begin{appendix}
\section{The derivation of (\ref{step0})}\label{app1}
In the calculation below we rely on the observation that the Gaussian process $Y(\cdot,x)$ with
exponentially decaying autocorrelation function should be well approximated
by a bounded process with a finite dependency range. In other words, 
we assume that there are a constants $C,\delta$: $Y^T(t,x)Y(t,x)\leq C$ for all $t,x$ and
the processes $(Y(t,x))_{t <T}$ and $(Y(t,x))_{t > T+\delta}$ are independent for any $T\in \R$
\footnote{It is the absence of error estimates associated with
this approximation which makes our present discussion non-rigorous.}.

To estimate a probability in terms of exponential moments, we follow
the logic of Chernoff bound (the exponential version of Chebyshev's inequality)
and notice the following elementary inequality: for any $\lambda \in \R$,
\beast
\ind(x \in (y\!-\!\eta, y\!+\!\eta))\!\leq\! e^{\lambda(x-y)+\eta |\lambda|}\ind(x \in (y-\eta, y+\eta)).
\eeast 
The corresponding $n$-dimensional generalisation is
\bea\label{chernoff}
\ind(x \in b_\eta(y))\leq e^{\lambda^T(x-y)+\eta n ||\lambda||_\infty}\ind(x \in b_\eta(y)),
\eea
where $\lambda,x,y \in \R^n$, $||\lambda||_\infty=\max_{1\leq \alpha \leq n}|\lambda_\alpha|$.
Using (\ref{chernoff}),
\begin{widetext}
\bea\label{stepa1}
\Prx \left[X(k \Delta t)\!\!\in\!\! b_\eta(x_k), 1\!\leq\! k\! \leq \!P \right]
\!\!=\!\!\E\!\!
\left[ \prod_{k=1}^P \!\!\ind\!\!\left(X(k\Delta t) \!\!\in\!\! b_\eta(x_k)\right) \right]
\!\!\leq\!\! \E \!\!\left[ e^{\sum_{l=1}^P\left[\frac{\lambda_l^T}{\epsilon} (X(l\Delta t)-x_l)+\eta n ||\lambda_l||_\infty\right]}\!\!
\prod_{k=1}^{P} \!\!\ind\!\!\left(X(k\Delta t)\!\! \in \!\!b_\eta (x_k)\right) \right]~~~~
\eea
\end{widetext}
Next we need to calculate $X(k\Delta t)-x_k$ for each $k$ by solving
(\ref{modeleq}) over the time interval $[(k-1)\Delta t, k\Delta t]$:
Denote the right hand side of the equation for $\dot{X}(\tau)$ by $f(\tau/\epsilon,X(\tau))$.
Our assumptions imply that both $||f||_\infty$ and $||\nabla_2 f||_\infty$ are bounded
by some constants, which will be denoted by $M_0$ and $M_1$ correspondingly.   
Expanding $f$ in Taylor series in the second argument one finds
\bea\label{ema}
\!\!\!\!\!\!\lambda^T \dot{X}(\tau)\!=\!\lambda^T\! f(\tau/\epsilon, x_{k-1})\!+\!\rho_k,
\tau \!\!\in\!\! [(k\!-\!1)\Delta t,k\Delta t],
\eea
where $\rho_k=\lambda^T\nabla_2 f(\tau/\epsilon,(1-c)x_{k-1}+cX(\tau))(X(\tau)-x_{k-1})$,
for some $c\in (0,1)$. Here we used the mean value form of the remainder for the
Taylor series. Using the bound $||\nabla_2 f||_\infty\leq M_1$
and noticing that $\nabla_2 f$ is an $n\times n$ matrix,
\bea\label{stepa2}
|\rho_k|\leq n^2 ||\lambda||_\infty M_1 ||X(\tau)-x_{k-1}||_\infty.
\eea
The estimate of the size of $X(\tau)-x_{k-1}$  uses the equation for
$X$ once more:
\beast
&&||X(\tau)-x_{k-1}||_\infty
\\
&&\leq\!\! ||X(\tau)\!-\!X(\Delta t(k-1))||_\infty\!\!+\!\!||X(\Delta t (k-1))\!-\!x_{k-1}||_\infty
\\
&&\leq \eta+|| \int_{(k-1)\Delta t}^\tau d\tau' f(\tau'/\epsilon, X(\tau'))||_\infty
\leq \eta+\Delta t M_0.
\eeast
The penultimate step uses the fact that the indicators under the sign of the expectation
in (\ref{stepa1}) enforce the constraint $||X((k-1)\Delta t)-x_{k-1}||_\infty<\eta$; the 
last step uses the bound $||f||_\infty<M_0$. Putting it all together we find that
\bea\label{bounda1}
|\rho_k|\leq n^2 ||\lambda||_\infty(\eta+\Delta t M_0)M_1.
\eea
Integrating (\ref{ema}) over the interval $[(k-1)\Delta t, k\Delta t]$ we conclude
that
\bea
&&\lambda_k^T (X(k\Delta t)-x_{k})=\lambda_k^T(x_{k-1}-x_k)\nonumber\\
&&+\lambda_k^T
\int_{(k-1)\Delta t}^{k\Delta t}\!\!\!\!\!\!\!\!\!\!\!\!d\tau f(\tau/\epsilon, x_{k-1})+\tilde{\rho}_k,
\label{bounda2}
\eea
where $||\tilde{\rho}_k||_\infty\leq ||\lambda_k||_\infty(n\eta+n^2(\eta+\Delta tM_0)M_1\Delta t)$.
Notice the extra contribution to the error term coming from one more application of
the bound $||X((k-1)\Delta t)-x_{k-1}||_\infty\leq \eta$.
Substituting (\ref{bounda2}) into (\ref{stepa1}) and upper-bounding
the product of the indicators by $1$, one arrives at the following intermediate result:
\bea\label{stepa10}
&&\Prx\left[X(k \Delta t)\in b_\eta(x_k), k = 1,\ldots, P \right]
\nonumber\\
&&\leq e^{\sum_{k=1}^P \frac{\lambda_k^T}{\epsilon}\left(x_{k-1}-x_{k}\right)}
\nonumber\\
&&\E
\left[ \prod_{k=1}^P e^{\frac{\lambda_k^T}{\epsilon} 
\left(\int_{(k-1)\Delta t}^{k\Delta t}d\tau f(\tau/\epsilon,x_{k-1})+R_k\right)}\right],
\eea
where $R_k=||\lambda_k||_\infty(2n\eta+n^2(\eta+\Delta tM_0)M_1\Delta t)$.
It remains to approximate the expectation in (\ref{stepa10}) by the product of expectations. To this end we write
\bea\label{stepa20}
 &&\lambda_k^T\int_{(k-1)\Delta t}^{k\Delta t}d\tau f(\tau/\epsilon,x_{k-1})=\epsilon\lambda_k^T \int_{(k-1)\Delta t/\epsilon}^{k\Delta t/\epsilon}\!\!\!\!d\tau f(\tau,x_{k-1})
 \nonumber\\
 &&=\epsilon\lambda_k^T \int_{(k-1)\Delta t/\epsilon+\delta}^{k\Delta t/\epsilon-\delta}d\tau f(\tau,x_{k-1})+E_k,~~~~~~~~~
\eea
where the bound $||f||_\infty\leq M_0$ implies that $|E_k|\leq 2\epsilon n ||\lambda_k||_\infty M_0\delta$. 
Crucially, notice that $f(\tau/\epsilon,x)$
depends on $Y(\tau/\epsilon,x)$ only. Therefore, the random variables $\epsilon\lambda_k^T \int_{(k-1)\Delta t/\epsilon-\delta}^{k\Delta t/\epsilon+\delta}d\tau f(\tau,x_{k-1})$,
$1\leq k\leq P$ are mutually independent due to the finite dependency
range $\delta$ of the process $Y$.  Substituting (\ref{stepa20}) into (\ref{stepa10}) 
and exploiting the independence one finds
\beast
&&\Prx\left[X(k \Delta t)\in b_\eta(x_k), k = 1,\ldots, P \right]
\nonumber\\
&&\leq e^{\sum_{k=1}^P \frac{\lambda_k^T}{\epsilon}\left(x_{k-1}-x_{k}+\tilde{R}_k\right)}
\prod_{k=1}^P \E
\left[e^{\lambda_k^T \int_{(k-1)\Delta t/\epsilon}^{k\Delta t/\epsilon}d\tau f(\tau,x_{k-1})}\right],
\eeast
where $\tilde{R}_k=||\lambda_k||_\infty(4\epsilon n M_0 \delta+2n\eta+n^2(\eta+\Delta tM_0)M_1\Delta t)$.
In the last expression we extended the integration interval back to $[(k-1)\Delta t,k\Delta t]/\epsilon$, which explains
the doubling of the $\delta$-dependent contribution to error term.
Finally, let us notice that the total error term
$R(\epsilon, \Delta t, \eta):=\sum_{k=1}^P \tilde{R}_k$ has the following property:
\bea
\lim_{\mu\rightarrow 0}\lim_{\Delta t\rightarrow 0}\lim_{\epsilon\rightarrow 0}R(\epsilon, \Delta t, \mu\Delta t)=0.
\eea
The derivation of (\ref{step0}) is complete.
\section{The derivation of (\ref{step12})}\label{app2}
First of all let us explain what we mean by a 'nice' set of functions $D$.
To this end,  we need to introduce one more notation. For
$f \in C([0,T],\R^n)$ let
\bea
&&B_\eta(f)\!\!=\!\!\{g\! \in \!C([0,t],\R^n):\!\!\!\!\!\!\!\!\!\!\! \sup_{\tau \in [0,t], 1\leq \alpha \leq n}\!\!\!\!\!\!\!\!\!\!\! 
|f_\alpha(\tau)\!\!-\!\!g_\alpha(\tau)|\!<\!\eta\}.~~
\eea
This is an infinite-dimensional generalization of the hypercube $b_\eta$ introduced above.
We say that the set $D$ is nice if for any $\eta>0$ we can find finitely many smooth functions 
$x^{(1)}, x^{(2)}, \ldots, x^{(M)} \in C([0,T],\R^n):$ 
\bea\label{appbincl}
D\subset \bigcup_{j=1}^M B_\eta(x^{(j)}).
\eea
In other words $D$ can be covered by finitely many hypercubes of any positive `linear size'
\footnote{A mathematician would say that $D$ is a totally bounded subset of the complete 
space $C([0,T],\R^n,||\cdot||_\infty)$ and is therefore compact.}.
Let $x^{(j)}_k=x^{(j)}(k\Delta t)$, $1\leq k\leq P$. Then
\begin{widetext}
\beast
&&\Prx[X \in D]\stackrel{(\ref{appbincl})}{\leq} \Prx\left[X\in \bigcup_{j=1}^M B_\eta (x^{(j)})\right]=\Prx[\exists j\leq M: X \in B_\eta(x^{(j)})]
=\Prx\left[\exists j\leq M: X(\tau)\in b_\eta\left(x^{(j)}(\tau)\right), \tau\in [0,t]\right]\\
&&\leq \Prx\left[\exists j\leq M: X(k\Delta t)\in b_\eta\left(x^{(j)}_k\right), 1\leq k\leq P\right]
\stackrel{(*)}{\leq} \sum_{k=1}^M \Prx\left[X(k\Delta t)\in b_\eta\left(x^{(j)}_k\right), 1\leq k\leq P\right]
\\
&&\leq M\max_{1\leq j\leq M}\Prx\left[\bigcap_{k=1}^P \left\{X(k\Delta t)\in b_\eta(x_k^{(j)}) \right\} \right].
\eeast
All of the above steps should be self-explanatory, let us just notice that the inequality $(*)$ is the union bound.
Taking the logarithm of both sides of the derived inequality and using the bound (\ref{step1})
one finds
\beast
\epsilon \Prx[X\in D]\leq \epsilon \log M+\epsilon  \max_{1\leq j\leq M}\log
\Prx\left[\bigcap_{k=1}^P \left\{X(k\Delta t)\in b_\eta(x_k^{(j)}) \right\} \right]
\leq \epsilon \log M+\epsilon  \sup_{x \in D}\log
\Prx\left[\bigcap_{k=1}^P \left\{X(k\Delta t)\in b_\eta(x_k) \right\} \right]
\\
\stackrel{(\ref{step1})}{\leq}\epsilon\log M+R+O(\epsilon P)
+\sup_{x \in D}\left[\sum_{p=1}^P  \Delta t
\lambda_p^T\left(\frac{x_{p-1}-x_{p}}{\Delta t}-\nu x_{p-1} \right)
-\frac{1}{2}\sum_{p=1}^P \Delta t \int_\R\frac{dk}{2\pi}\log 
\det\left(I-2\lambda_p^TM\tilde{C}(k,x_{p-1})\right)\right].
\eeast
As a result,
\bea \label{step11}
\limsup_{\epsilon \rightarrow 0} \epsilon \Prx[X \in D]\leq \lim_{\epsilon \rightarrow 0}R\!+\!
\sup_{x \in D}
\sum_{p=1}^P  \Delta t\left(
\lambda_p^T\left(\frac{x_{p-1}\!-\!x_{p}}{\Delta t}\!-\!\nu x_{p-1} \right)
\!-\!\frac{1}{2} \int_\R\frac{dk}{2\pi}\log 
\det\left(I\!-\!2\lambda_p^TM\tilde{C}(k,x_{p-1})\right)\right).~~
\eea
\end{widetext}
Finally, notice that the left-hand-side of (\ref{step11}) does not depend on $\Delta t, \eta$.  
Let $\lambda,x$ be a pair of $\R^n$-valued functions on $[0,t]$ such that 
$$ 
\lambda(k\Delta t)=\lambda_k,~x(k\Delta t)=x_k, 1\leq k\leq P.
$$
Setting $\eta=\mu \Delta t$, applying $\lim_{\mu\rightarrow 0}\lim_{\Delta t\rightarrow 0}$ to both
sides of (\ref{step11}) and using the property (\ref{errest}) of the error term one arrives
at (\ref{step12}).
\section{Lower bound on $\log \Prx[X\in D]$}\label{appLB}
For the lower bound, let us take the pair $x,\lambda \in C([0,t], \R^n)$ to be 
the solution to the Euler-Lagrange 
equations describing the critical points of (\ref{seff}) and assume that the solution
is unique and smooth. The boundedness of the right hand side of
the equation for $\dot{x}$  
means that there is $M_2>0$ such that $||\dot{x}(t)||_\infty<M_2$ for 
all $t\in [0,t]$. 

Let us fix $\Delta t>0$.
Using the above  bound on $\dot{x}$ and the bound $||f||_\infty<M_0$ discussed
in the text above (\ref{ema}), it is easy to establish the following: if $X(t)\in b_\eta(x(t))$
at some time $t$, then for all $\tau \in [t, t+\Delta t]$,
\bea\label{lbgl1}
||X(\tau)-x(\tau)||_\infty \leq \eta+(M_0+M_2)\Delta t.
\eea
Let $x_{k}=x(k\Delta t), \lambda_k=\lambda(k\Delta t)$, $1\leq k\leq P$,
where $P=\lfloor \frac{t}{\Delta t}\rfloor$. Choose $\rho>0$: $B_\rho(x)\subset D$. Then
\bea\label{lbgl2}
\Prx[X\in D]\geq \Prx[X\in B_\rho(x)]\\
\nonumber
\geq 
\Prx[X(k\Delta t)\in b_\eta(x_k), 1\leq k\leq P]
\eea
provided $\eta>0$ and $\Delta t>0$ are such that $\eta+(M_0+M_2)\Delta t<\rho$:
given such a choice, the estimate (\ref{lbgl1}) implies the inclusion of events
$\cap_{k=1}^P \{X(k\Delta t)\in b_\eta(x_k)\}\subset \{X\in B_\rho(x)\}$,
which leads to the claimed inequality in (\ref{lbgl2}).


Following the steps which led to (\ref{bounda2}), one finds
\bea
\!\!\!\!\!\!\!X_k\!\!-\!\!X_{k-1}\!\!=\!\!\int_{(k-1)\Delta t}^{k\Delta t}
\!\!\!\!\!\!\!\!\!\!\!\!f(\tau/\epsilon,x_{k-1})d\tau+V_k,
\label{boundc1}
\eea
where $X_k:=X(k\Delta t)$ and $||V_k||_\infty\leq nM_1(\eta+\Delta t M_0)\Delta t$.
Let 
\beast
F^{(\epsilon)}_k=\int_{(k-1)\Delta t+\epsilon \delta}^{k\Delta t}f(\tau/\epsilon,x_{k-1})d\tau.
\eeast
By the finite dependency assumption the random variables $\left(F^{(\epsilon)}_k\right)_{k\geq 1}$ are independent. Define $F_{k-1}:=F_{k-1}^{(0)}$.
Then right hand side of (\ref{boundc1}) is equal to $F_{k-1}+V_k$.
Notice the following elementary inequality:
\bea
\ind (X+v\in b_\eta(x))\geq \ind(X\in b_{\eta-||v||_\infty}(x)).\label{boundc4}
\eea
The right hand side is non-zero provided $||v ||_\infty <\eta$. 
The following estimate is based on (\ref{lbgl2}), the independence of $(F_k^{(\epsilon)})_{k\geq 1}$,
the inequality (\ref{boundc4}) and the tower property of conditional probabilities:
 \begin{widetext}
\bea
\Prx[X\in D]\geq \Prx[X\in B_\rho(x)]
\geq  \E[\prod_{k=1}^P\ind(X_k\in b_\eta(x_k))]=\E\left[\E\left[\prod_{k=1}^P\ind(X_k\in b_\eta(x_k))\bigg\vert
(Y_\tau)_{0\leq \tau\leq (P-1)\Delta t/\epsilon}  \right]\right]\nonumber\\
=\E\left[\prod_{k=1}^{P-1}\ind(X_k\in b_\eta(x_k))
\E\left[\ind(X_P\in b_\eta(x_P))\bigg\vert
(Y_\tau)_{0\leq \tau\leq (P-1)\Delta t/\epsilon}  \right]\right]
=\E\left[\prod_{k=1}^{P-1}\ind(X_k\in b_\eta(x_k))\right.\nonumber\\
\left.\E\left[\ind(X_{P-1}+F_{P}^{(\epsilon)}+v\in b_\eta(x_P))\bigg\vert
(Y_\tau)_{0\leq \tau\leq (P-1)\Delta t/\epsilon}  \right]\right]
\geq \min_{y_{P-1}\in b_{\eta}(x_{P-1})}
\E\left[\prod_{k=1}^{P-1}\ind(X_k\in b_\eta(x_k))\right.\nonumber\\
\left.\E\left[\ind(y_{P-1}+F_{P}^{(\epsilon)}+v\in b_\eta(x_P))\bigg\vert
(Y_\tau)_{0\leq \tau\leq (P-1)\Delta t/\epsilon}  \right]\right]
= \min_{y_{P-1}\in b_{\eta}(x_{P-1})}
\E\left[\prod_{k=1}^{P-1}\ind(X_k\in b_\eta(x_k))\right]\nonumber\\
\E\left[\ind(y_{P-1}+F_{P}^{(\epsilon)}+v\in b_\eta(x_P)) \right]
\geq \prod_{k=1}^P\min_{y_{k-1}\in b_{\eta}(x_{k-1})}
\E\left[\ind(y_{k-1}+F_{k}^{(\epsilon)}+v\in b_\eta(x_k)) \right]\nonumber\\
=
\prod_{k=1}^P\min_{y_{k-1}\in b_{\eta}(x_{k-1})}
\E\left[\ind(y_{k-1}+F_{k}+w\in b_\eta(x_k)) \right]
\geq \prod_{k=1}^P\min_{y_{k-1}\in b_{\eta}(x_{k-1})}
\E\left[\ind(y_{k-1}+F_{k}\in b_{\eta-r}(x_k)) \right]
\label{stepc100}
\eea
\end{widetext}
Here $v$ and $w$ are a shorthand notation for random errors satisfying deterministic bounds on their
norms, $||v||_\infty\leq  nM_1(\eta+\Delta t M_0)\Delta t+\epsilon M_0 \delta$,
$||w||_\infty\leq  nM_1(\eta+\Delta t M_0)\Delta t+2\epsilon M_0 \delta$, and
$r= nM_1(\eta+\Delta t M_0)\Delta t+2\epsilon M_0 \delta$.
The following steps are standard in the context of the theory of large deviations: let $\eta'=\eta-r$. Then
notice that
\begin{widetext}
\bea
&\E&\left[ \ind\left(F_k\in b_{\eta}(x_k-y_{k-1})\right) \right]
=\E\left[ e^{-\frac{1}{\epsilon}\lambda_k^TF_k}e^{\frac{1}{\epsilon}\lambda_k^TF_k}
\ind\left(F_k\in b_{\eta'}(x_k-y_{k-1})\right) \right]
\geq e^{-\frac{\lambda_k(x_k-y_{k-1})+n||\lambda_k||_\infty\eta'}{\epsilon}}
\nonumber\\
&\E&\left[e^{\frac{1}{\epsilon}\lambda_k^TF_k}\ind\left(F_k\in b_{\eta'}(x_k-y_{k-1})\right) \right]=
e^{-\frac{\lambda_k(x_k-y_{k-1})+n||\lambda_k||_\infty}{\epsilon}}
\E\left[ e^{\frac{1}{\epsilon}\lambda_k^TF_k}\right]
\E^{(\lambda_k)}\left[\ind\left(F_k\in b_{\eta'}(x_k-y_{k-1})\right) \right],
\label{stepc1}
\eea 
\end{widetext}
where 
\beast
\E^{(\lambda_k)}[\bullet ]:=\frac{\E[e^{\frac{1}{\epsilon}\lambda_k^TF_k} \bullet]}{\E\left[ e^{\frac{1}{\epsilon}\lambda_k^TF_k}\right]},
\eeast
is the expectation with respect to the probability measure tilted by the exponential 
factor $e^{\frac{1}{\epsilon}\lambda_k^TF_k}$. 
The derivation of (\ref{stepc100},\ref{stepc1}) did not use any assumptions about the 
sequence $(\lambda_1, \ldots \lambda_P)$. 
Now let us choose the sequence in such a way that
\bea\label{stepc2}
\lim_{\epsilon\rightarrow 0}\E^{(\lambda_k)}[F_k]=x_k-y_{k-1},~1\leq k\leq P,
\eea
which coincides with the discretised version of the Euler-Lagrange equations $\delta S_{eff}/\delta \lambda(\tau)=0,~0<\tau<t$ if $y_k=x_k$ for all $k$'s.
 Equivalently, 
 \bea\label{stepc3}
\!\!\!\!\frac{\partial}{\partial \lambda_k}\lim_{\epsilon\rightarrow 0}\!\epsilon\! \log \E[e^{\frac{1}{\epsilon}\lambda_k^TF_k}]
\!=\!x_k\!-\!y_{k-1},1\leq k\leq P.
\eea
Recall that an explicit formula derived with the help of  Widom's theorem shows that
$\lim_{\epsilon \rightarrow 0}\epsilon \log \E[e^{\frac{1}{\epsilon}\lambda_k^TF_k}]$
is finite, see (\ref{fdet1}) and (\ref{fdet2}).  
Calculating the second $\lambda$-derivative of $\log\E[e^{\frac{1}{\epsilon}\lambda^TF_k}]$
one finds that 
\bea\label{stepc4}
\hspace{-0.7cm}
\lim_{\epsilon\rightarrow 0}\mbox{Cov}^{(\lambda_k)}[ F_k]=\lim_{\epsilon\rightarrow 0}
\epsilon^2  
\partial_{\lambda}\otimes\partial_{\lambda}\log\E[e^{\frac{1}{\epsilon}\lambda^TF_k}]=0.
\eea
Expressions (\ref{stepc2}), (\ref{stepc4}) and Chebyshev's inequality imply that
\beast
\lim_{\epsilon \rightarrow 0}\E^{(\lambda_k)}\left[\ind\left(F_k\in b_{\eta'}(x_k-y_{k-1})\right) \right]=1.
\eeast
Using this observation in (\ref{stepc100}, \ref{stepc1}) one finds that
\bea\label{boundc7}
&&\liminf_{\epsilon\rightarrow 0}\epsilon \log\Prx[X\in D]
\nonumber\\
&\geq& \sum_{k=1}^P \min_{y_{k-1}\in b_{\eta}(x_{k-1})}\left\{\left(\lambda_k^T (y_{k-1}-x_k)-n||\lambda_k||_\infty \eta''\right)
\right. \nonumber\\
&+&\left.\lim_{\epsilon \rightarrow 0}\epsilon\log
\E\left[ e^{\frac{1}{\epsilon}\lambda_k^TF_k}\right]\right\},
\eea
provided the sequence $(\lambda_k)$ solves (\ref{stepc2}). 
Here 
$$\eta''=\eta'\mid_{\epsilon=0}=\eta-r\mid_{\epsilon=0}=\eta-nM_1(\eta+\Delta t M_0)\Delta t.$$

As the right hand side of (\ref{boundc7}) does not depend on $\eta$ and $\Delta t$,
one can set $\eta=a(\Delta t)^\mu$, where $a>0$ and $\mu>1$ and take the limit 
$\Delta t\rightarrow 0$. In the limit, using that $y_k\in b_{\eta}(x_k)$, one finds that the system of equations (\ref{stepc3}) becomes
\[
\!\!\!\!\frac{\partial}{\partial \lambda(\tau)}\lim_{\Delta t,\epsilon\rightarrow 0}\!\frac{\epsilon}{\Delta t}\! \log \E[e^{\frac{1}{\epsilon}\lambda_k^TF_k}]\mid_{k=\frac{\tau}{\Delta t}}
\!=\!\dot{x}(\tau),~\tau\in[0,t].
\]
Due to (\ref{fdet1}), (\ref{fdet2}), the above equation coincides with the 
Euler-Lagrange equation $\frac{\delta S_{eff}}{\delta \lambda(\tau)}=0$,
where $S_{eff}$ is given by (\ref{seff}). The bound (\ref{boundc7}) becomes
\bea\label{boundc9}
\!\!\!\!\!\liminf_{\epsilon \rightarrow 0}\epsilon \log\Prx[X\in D]\geq 
\int_0^t d\tau \bigg(-\lambda^T(\tau)\cdot{x}(\tau)
\\
\nonumber
+\lim_{\Delta t\rightarrow 0}\lim_{\epsilon \rightarrow 0}\frac{\epsilon}{\Delta t} 
\log
\E\left[ e^{\frac{1}{\epsilon}\lambda_k^TF_k}\right]\big|_{k\Delta t=\tau}\bigg),
\eea
where the functions $\lambda, x$ solve the Euler-Lagrange equations associated with 
the effective action functional $S_{eff}$.
Furthermore,  the right hand side
of (\ref{boundc9}) coincides with the effective action functional (\ref{seff}). Therefore,
by the assumed uniqueness of the solution to
the Euler Lagrange equaitons, the right
hand side of (\ref{boundc9}) must coincide with (\ref{step13}). The lower bound is derived.
\section{Widom's theorem}\label{app3}
To the best of our knowledge, a proof 
of Widom's theorem has never been published. In \cite{widom} Widom simply 
formulates the theorem, states that it can be easily verified by taking the continuous
limit of the corresponding statement for large Toeplitz matrices and then 
moves on to the main topic of the paper: the asymptotic of Fredholm determinants for operators acting on spaces of functions of several variables.
Thus there is a gap in the story, which we partially fill in the present Appendix by deriving (\ref{fdet2}).
In our proof we use the probabilistic method developed in the original paper by Kac \cite{kac}.\\
\\
{\em {\bf Theorem.}
Let $K: \R\rightarrow \R^{N\times N}$ be an $N\times N$ matrix-valued function of one variable.
Assume that $K$ is even ($K(t)=K(-t)$, for any $t \in \R$) and non-negative ($K_{ij}(t)\geq 0$ for any $t \in \R$ and $1 \leq i,j\leq N$).
Assume in addition that
\bea
&&\int_\R |t|K(t)dt<\infty,\label{converge}\\
&&\int_\R \sum_{k=1}^N K_{ki}\leq 1,~ 1\leq i \leq N.\label{norm} 
\eea
The function $K$ can be regarded as a kernel of an integral operator $\hat{K}$ acting on square-integrable functions from $\R$ to $\R^N$,
\bea
f\mapsto \hat{K}f(t)=\int_\R d\tau K(t-\tau)f(\tau),~ t \in \R.
\eea 
Then there is $\lambda_{max}>0$ such that for any $\lambda: |\lambda|<\lambda_{max}$ the Fredholm determinant $\text{Det}(I-\lambda \hat{K}_T)$ exists and
\bea
\log\text{\text{Det}}(I-\lambda \hat{K}_T)=\!\!T\!\!\int_\R \frac{dk}{2\pi}\log\det(1-\lambda \tilde{K}(k))
\nonumber\\
+O(T^0),
\label{wta}
\eea
 where $\hat{K}_T$ is the restriction of $\hat{K}$ to functions on $[0,T]$ and
 \bea
\tilde{K}(k)=\int_\R dx e^{-ikx} K(x),~ k \in \R.
\eea
}

Let us sketch the proof of the theorem using, as we already mentioned, the probabilistic method used in \cite{kac} to prove a continuous version of Szegö's formula for the asymptotics of Toeplitz determinants. 
For a sufficiently small $|\lambda|$ we can calculate the Fredholm determinant using the trace-log formula,
\bea\label{trloga}
\log \text{Det}(I-\lambda \hat{K}_T)=-\sum_{n=1}^\infty \frac{1}{n}\lambda^n \text{Tr} \hat{K}_T^n,
\eea
where 
\beast
&&\text{Tr} \hat{K}_T^n=\int_{[0,T]^n}dx_1 dx_2\ldots dx_n 
\\
&&\text{tr} K(x_1-x_2)K(x_2-x_3)\ldots K(x_n-x_1).
\eeast
Using the cyclic property of trace and the fact that the function $K$ is even, we find
\bea
\frac{d}{dT}\text{Tr} \hat{K}_T^n=n\int_{[0,T]^n}dx_2dx_3\ldots dx_{n} 
\nonumber\\
\text{tr} K(-x_2)K(x_2-x_3)\ldots K(x_{n-1}-x_n)K(x_n).
\label{trder}
\eea
Consider the following discrete time Markov chain $\{X_n,S_n\}_{n\geq 0}$ on the state space $\R\times \{1,2,\ldots,N\}$:
\begin{enumerate}
\item $(X_0, S_0) \sim (\delta_0,U_N)$, 
where $U_N$ is the uniform distribution on $ \{1,2,\ldots,N\}$.
\item At each time step, the transition $(x,i)\rightarrow (y,k)$ happens with probability $K_{ki}(y-x)dy$. 
\end{enumerate} 
Notice that this is a Markov chain with killing, the survival probability when transitioning from state $(x,i)$ is $g_i(x):=\sum_{k=1}^N \int_\R K_{ki}(y-x)dy\leq 1$.
Examining the expression (\ref{trder}) for the derivative of the trace of the $n$-th power of $\hat{K}$, we see that it can be interpreted as the following expectation
with respect to the law of the chain $\{X_n,S_n\}_{n\geq 0}$:
\bea\label{trmc}
\!\!\!\!\!\!
\frac{d}{dT}\text{Tr} \hat{K}_T^n\!=\!Nn\E\left(\ind(X_n \in d0) \ind(S_n\!=\!S_0)\ind(\tau\!=\!n)  \right),
\eea
where $\tau$ is the first exit time of the chain from the interval $(0,T)\times \{1,2,\ldots,N\}$. To derive
the above expression we exploited the identity  $\ind(X_n \in d0)\ind(\tau \geq n) = \ind(X_n \in d0)\ind(\tau=n)$. 
Substituting (\ref{trmc}) into (\ref{trder}) and then (\ref{trloga}), we find
that
\beast
&&\frac{d}{dT}\log \text{Det}(I\!-\!\lambda \hat{K}_T)\!=\!-N\E\left(\lambda^{\tau}\ind(X_\tau \in d0)\ind(S_\tau=S_0) \right)\\
&&=-N\E\left(\lambda^{\tau_0}\ind(X_{\tau_0} \in d0)\ind(M_{\tau_0}<T)\ind(S_{\tau_0}=S_0) \right),
\eeast
where $\tau_0$ is the first exit time from $(0,\infty)\times\{1,2,\ldots, N\}$, $M_{\tau_0}=\max_{1\leq n<\tau_0}(X_n)$.
As $\log\det(I-\lambda\hat{K}_0)=0$, we can integrate the last expression to find
\beast
&&\log\text{Det}(I-\lambda\hat{K}_T)
\\
&&=-N\E\left(\lambda^{\tau_0}\ind(X_{\tau_0} \in d0)(T-M_{\tau_0})_+\ind(S_{\tau_0}=S_0)  \right),
\eeast
where $(x)_{+}:=\max(x,0)$. Noticing that $T-(T-M)_{+}=\min(T,M)$, we can re-arrange the above expression as follows:
\beast
&&\log\text{Det}(I-\hat{K}_T)=-NT\E\left(\lambda^{\tau_0}\ind(X_{\tau_0} \in d0)\ind(S_{\tau_0}=S_0)  \right)
\nonumber\\
&&+N\E\left(\lambda^{\tau_0}\ind(X_{\tau_0} \in d0)\min(T,M_{\tau_0})\ind(S_{\tau_0}=S_0)  \right),
\eeast
This is an exact expression for the Fredholm determinant as an expectation with respect to the law of the Markov chain we defined.
In many cases it allows for an efficient computation of the large-$T$ expansion of the Fredholm determinant using purely probabilistic methods. 
For us it is sufficient to check that $\lim_{T\rightarrow \infty} \min(T,M_{\tau_0})=M_{\tau_0}$, which implies that
\bea
\log\text{Det}(I\!-\!\lambda\hat{K}_T)&=&\!-NT\E\left(\lambda^{\tau_0}\!\ind(X_{\tau_0}\! \in\! d0)\ind(S_{\tau_0}=S_0)  \right)
\nonumber\\
&+&O(T^{0}).
\label{penal}
\eea
To calculate the expectation entering the leading term we use the following combinatorial lemma (see e. g. \cite{feller}, volume $2$):
Let $(0,R_1,R_1+R_2, \ldots, R_1+R_2+\ldots+R_{n-1},0)$ be the first $n$ $\R$-projections of the states of the chain with $\tau_0=n$. 
Then
 \bea
&&\sum_{p=0}^{n-1} \prod_{k=1}^{n-1}\ind(R_{1+p}+R_{2+p}+\ldots+R_{k+p}>0)
\nonumber\\
&&=1~\text{a. s.},
~0\leq p\leq n-1.
\label{cl}
\eea
The addition of subscripts in the above formula should be understood modulo $n$. The above statement is very general and relies
only on the absence of atoms in the transition probabilities $K(y-x)dy$. 

In this case, for any sequence $(0,R_1,R_1+R_2, \ldots, R_1+R_2+\ldots+R_{n-1},0)$,
its graph
will almost surely have a unique global minimum, so there will be a unique cyclic permutation $(0,R_{1+p},R_{1+p}+R_{2+p}, \ldots, R_{1+p}+R_{2+p}+\ldots+R_{n-1+p},0)$,
whose graph will stay positive between times $1$ and $n-1$. Then 
\begin{widetext}
\bea
&&N\E\left(\lambda^{\tau_0}\ind(X_{\tau_0} \in d0)\ind(S_{\tau_0}=S_0)  \right)
=N\sum_{n=1}^\infty \lambda^n\E\left(\ind(X_{\tau_0} \in d0)\ind(S_{\tau_0}=S_0) \ind(\tau_0=n) \right)
\nonumber\\
&=&\sum_{n=1}^\infty \lambda^{n} \int_{\R^n} dr_1 \ldots dr_n \text{tr}(K(r_1)\ldots K(r_n))\delta(r_1+\ldots +r_n)\prod_{k=1}^{n-1}\ind(r_1+\ldots+r_k>0)
\nonumber\\
&=&
\sum_{n=1}^\infty \frac{\lambda^{n}}{n}\int_{\R^n} dr_1 \ldots dr_n \text{tr}(K(r_1)\ldots K(r_n))\delta(r_1+\ldots +r_n)\sum_{p=0}^{n-1}\prod_{k=1}^{n-1}\ind(r_{1+p}+\ldots+r_{k+p}>0)
\nonumber\\
&=&
\sum_{n=1}^\infty \frac{\lambda^n}{n}\int_{\R^n} dr_1 \ldots dr_n \text{tr}(K(r_1)\ldots K(r_n))\delta(r_1+\ldots +r_n)
\nonumber\\
&=&
\sum_{n=1}^\infty \frac{\lambda^n}{n}\int_\R \frac{dk}{2\pi} \int_{\R^n} dr_1 \ldots dr_n e^{-ik(r_1+\ldots+r_n)}\text{tr}(K(r_1)\ldots K(r_n))
\nonumber\\
&=&
\sum_{n=1}^\infty \frac{\lambda^n}{n}\int_\R \frac{dk}{2\pi} \text{tr}(\tilde{K}(k_1)\ldots \tilde{K}(k_n))
=-\int_\R \frac{dk}{2\pi}\log \det (I-\lambda\tilde{K}(k)).
\label{finally}
\eea
\end{widetext}
The third inequality is the symmetrisation of the integrand with respect to all cycling permutations, the fourth
inequality is due to the combinatorial lemma (\ref{cl}).
Substituting (\ref{finally}) into (\ref{penal}), we arrive at the statement (\ref{wta}) of Widom's theorem.\\
\\
{\bf Remarks.} 
\begin{enumerate}
\item In \cite{widom} Widom presents a stronger version of the above statement which characterises the $O(T^0)$ term fully. For the current paper we only need the leading term.
\item The actual statement of Widom's theorem does not require the positivity of the kernel. In fact, all steps of the proof
presented below go through for signed kernels as well, but the probabilistic intuition guiding these steps is lost. See also
\cite{kac} for similar remarks about the original proof of Szeg\H{o}'s theorem by Marc Kac.
\item It is possible to give an alternative derivation of (\ref{fdet2}) 
based on the re-summation of the cumulant expansion for the expectation of
 a quadratic function of a Gaussian process. The downside of such
a derivation is difficulty in controlling the sub-leading terms. 
\end{enumerate}
\end{appendix}
\bibliography{quadev_revision2}
\end{document}